\documentclass[iop]{emulateapj}
\usepackage{verbatim}
\usepackage{gensymb}
\usepackage{booktabs}
\usepackage{array}
\usepackage{threeparttable}
\newcommand{\otoprule}{\midrule[\heavyrulewidth]}
\newcolumntype{+}{>{\global\let\currentrowstyle\relax}}
\newcolumntype{^}{>{\currentrowstyle}}

\begin{document}

\title{Investigating the Minimum Energy Principle in Searches for New Molecular Species - the Case of H$_{2}$C$_{3}$O Isomers}


\author{
Ryan A. Loomis\altaffilmark{1,2},
Brett A. McGuire\altaffilmark{3,4},
Christopher Shingledecker\altaffilmark{5},
Andrew Burkhardt\altaffilmark{5},
Chelen H. Johnson\altaffilmark{6},
Samantha Blair\altaffilmark{7},
Amy Robertson\altaffilmark{8},
and
Anthony J. Remijan\altaffilmark{3}}

\altaffiltext{1}{Corresponding author: rloomis@cfa.harvard.edu}
\altaffiltext{2}{Department of Astronomy, Harvard University, Cambridge, MA 02138}
\altaffiltext{3}{National Radio Astronomy Observatory, Charlottesville, VA 22904}
\altaffiltext{4}{Division of Chemistry and Chemical Engineering, California Institute of Technology, Pasadena, CA 91125}
\altaffiltext{5}{Department of Astronomy, University of Virginia, Charlottesville, VA 22904}
\altaffiltext{6}{Breck School, Golden Valley, MN 55422}
\altaffiltext{7}{Department of Natural Sciences, Dalton State, Dalton, GA 30720}
\altaffiltext{8}{Department of Astronomy, University of Arizona, Tucson, AZ 85721}

\begin{abstract}
Recently, \cite{Lattelais_et_al_2009} have interpreted aggregated observations of molecular isomers to suggest that there exists a ``minimum energy principle'', such that molecular formation will favor more stable molecular isomers for thermodynamic reasons.  To test the predictive power of this principle, we have fully characterized the spectra of the three isomers of C$_{3}$H$_{2}$O toward the well known molecular region Sgr B2(N). Evidence for the detection of the isomers cyclopropenone (c-C$_{3}$H$_{2}$O) and propynal (HCCCHO) is presented, along with evidence for the non-detection of the lowest zero-point energy isomer, propadienone (CH$_2$CCO). We interpret these observations as evidence that chemical formation pathways, which may be under kinetic control, have a more pronounced effect on final isomer abundances than thermodynamic effects such as the minimum energy principle.
\end{abstract}


\keywords{astrochemistry, ISM molecules, ISM: abundances, ISM: clouds, molecular processes}

\section{Introduction}

A long-standing goal of astrochemistry is to understand the physical and chemical evolution of interstellar sources through molecular observations. The formation and destruction of complex organic molecules are affected by the source conditions; observed molecular abundance ratios preserve information about the physical evolution of the region, acting as a chemical fingerprint. Extracting this information has proven to be exceedingly difficult, however, for even the simplest sources. Due to complex source structure, incomplete molecular inventories of sources, vast networks of molecular interactions, and only a partial understanding of interstellar chemical formation pathways, it is challenging to create models sufficiently accurate to predict observations \citep{Quan_&_Herbst_2007}. 

Numerous efforts have sought to address these issues by establishing complete molecular inventories for molecule-rich sources and characterizing basic conditions and abundances \citep[e.g.][]{Neill_et_al_2014, Crockett_et_al_2014}, and creating complete reaction networks with accurate reaction dynamics \citep[e.g.][]{Garrod_et_al_2008, Garrod_2013}. Still, the low densities, low temperatures, and long time scales of molecular processes in the interstellar medium (ISM) present a significant challenge to relating laboratory reaction dynamics and mechanisms to those occurring in the ISM \citep{Laas_et_al_2013}. Whether the reactions governing the formation of complex molecules are primarily under kinetic or thermodynamic control is of particular interest, as this significantly affects molecular abundance ratios and the formation of prebiotically relevant molecules \citep{Garrod_et_al_2008, Lovas_et_al_2010}.

Observations of molecular isomers provide a unique tool for directly probing hypotheses about reaction mechanisms in the ISM, as their formation and destruction routes directly influence their abundance ratios \citep{Hollis_et_al_2006}. While this is true for all molecules, isomers are particularly useful as they present a similar level of chemical complexity, and have the same constituent atoms, simplifying the problem tremendously.  This approach has historically proven to be fruitful; hypotheses of the formation chemistry of HCN led to the detection of its isomer, HNC \citep{Snyder_&_Buhl_1971}, and comparisons between the abundances and distributions of conformational isomers of methyl formate, ethanol, and vinyl alcohol have yielded important information about their formation pathways \citep{Neill_et_al_2012, Pearson_et_al_1997, Turner_&_Apponi_2001}.   Thus, understanding the root causes of molecular abundance ratios and accurately predicting which isomers are most likely to be observed is a highly desirable goal.  Comparing abundance ratios with the conversion barriers and zero-point energy differences between the different isomers can elucidate whether a thermal process or kinetically controlled formation mechanism is primarily responsible in determining the ratio.

Recently, \citep{Lattelais_et_al_2011, Lattelais_et_al_2010, Lattelais_et_al_2009} investigated the relative abundances of 32 detected molecular isomers and suggested that there exists a ``minimum energy principle" (MEP), generally applicable across a number of astronomical environments.  According to this principle, the isomer with the lowest zero-point bonding energy (i.e., the most energetically stable isomer) should be the most abundant, and thus most easily detected. \cite{Lattelais_et_al_2009} include a wide variety of sources in this analysis, including molecular clouds, hot cores/corinos, photodissociation regions, and asymptotic giant branch stars, and do not note any differences in trends between these sources.  Importantly, they make no mention of attempting to exclude sources where LTE assumptions would be unlikely to hold, and instead assert that LTE effects may be general across all sources used in their analysis. C$_4$H$_3$N and C$_2$H$_4$O$_2$ are noted as exceptions to the principle, but are dismissed by calling the abundance measurements of the former into question and proposing differential adsorption rates onto grains for the latter \citep{Lattelais_et_al_2010, Lattelais_et_al_2009}. However, studies of other common isomers (e.g. HCN/HNC) have shown that abundances and emission strength are dependent on the local environment, suggesting that there must be more to the picture than bonding energy ratios (see e.g. \cite{Sarrasin_et_al_2010} and references therein). We therefore assert that the chemical formation route and physical conditions of these species, especially complex organics, are more significant in establishing relative abundance ratios than the zero-point bonding energies, and thus the MEP is not generally applicable as proposed by Lattelais et al.

Toward this end, we have fully characterized the spectra of the three isomers of C$_{3}$H$_{2}$O toward the well-studied molecular region Sgr B2(N). The relative bonding energies of these isomers are shown in Figure 1.  Based on these bonding energies and with no other prior knowledge, the MEP would predict the most stable isomer, propadienone (CH$_{2}$CCO), to be observed as the most abundant of the three isomers \citep{Karton_&_Talbi_2014}. However, propadienone has never been detected toward an astronomical source, while both propynal and cyclopropenone have been detected in molecular clouds \citep{Irvine_et_al_1988, Hollis_et_al_2006}.  We present evidence for the detection of the two higher energy isomers cyclopropenone (c-C$_{3}$H$_{2}$O) and propynal (HCCCHO), along with evidence of the non-detection of propadienone toward Sgr B2(N). We interpret this to show that chemical formation pathways, which may be under kinetic control, have a more pronounced effect on final isomer abundances than thermodynamic effects such as the MEP.

\begin{figure}[h!]
\centering
\includegraphics[scale=.25]{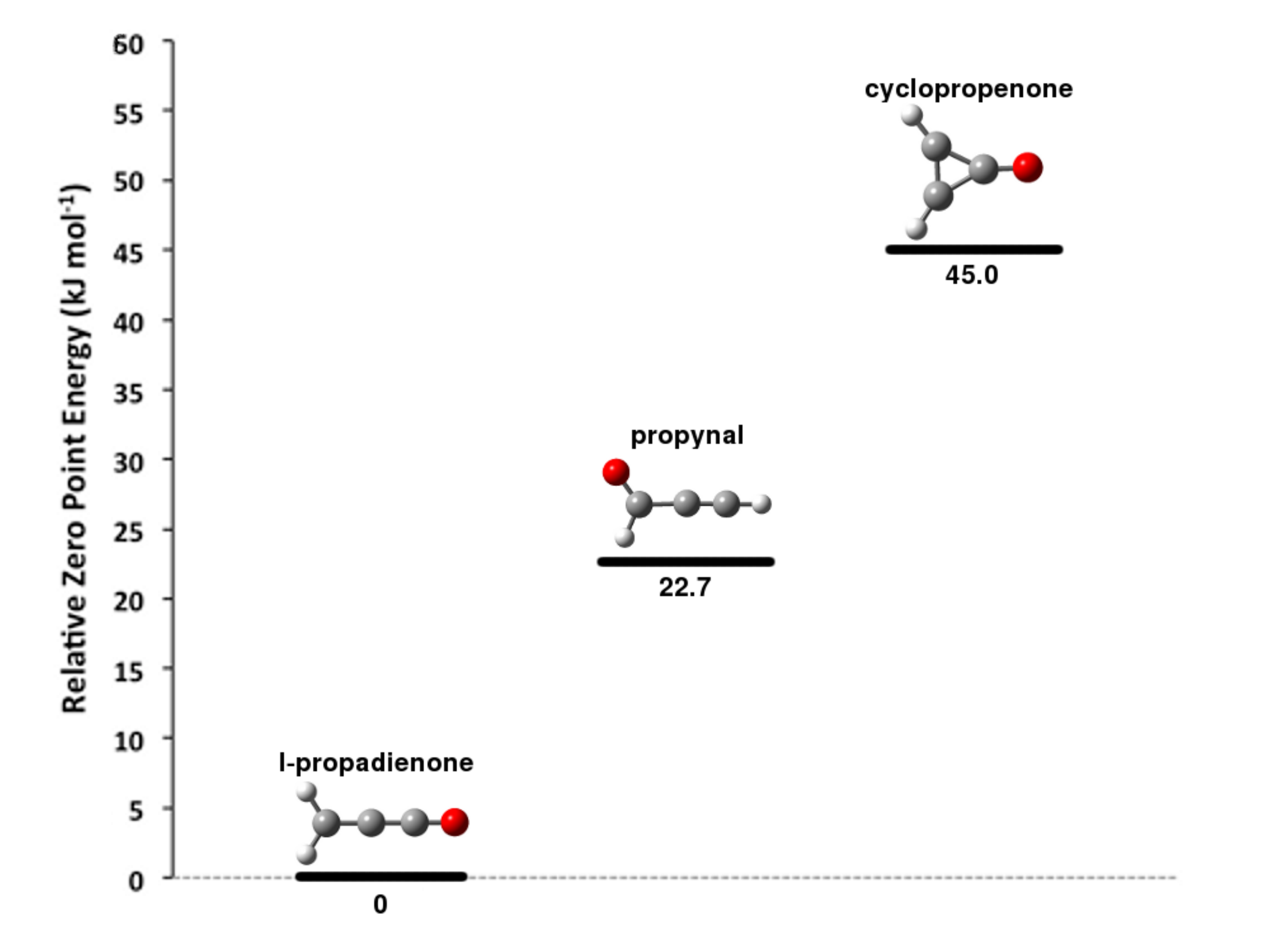}
\caption{{\small Relative binding energies of the three C$_{3}$H$_{2}$O isomers}}
\end{figure}

\section{Observations}

Sgr B2(N) is one of the preeminent sources for complex molecule detections, and the majority of all molecules used in the analysis of \cite{Lattelais_et_al_2009} have been detected there, thus it serves as a reasonable source to test the predictions of the MEP.  Sgr B2(N) has a heterogenous temperature and density structure, with both a hot core and a colder, less dense envelope, yielding multiple velocity components and a complex chemistry \citep{Belloche_et_al_2013}.  For a full discussion of the effect of this structure on molecular inventories of the respective regions, see \cite{Neill_et_al_2014}.  Significantly, many of the large complex organic molecules observed toward Sgr B2(N) (including some traditionally considered to be hot-core molecules) have been seen to have sub-thermal emission, with significant non-LTE effects \citep[and references therein]{Neill_et_al_2012, Loomis_et_al_2013, Zaleski_et_al_2013}.  These molecules have very low rotational temperatures, suggesting that they are extended and belong to the envelope, but the non-LTE effects make determinations of column density and temperature extremely difficult without collisional cross-sections \citep{Faure_et_al_2014}.  As all three of the molecules searched for in this study would be expected to originate in colder environments, these non-LTE effects are likely very relevant, and are discussed in Section 4.1.

All known strong (predicted to be above the measured noise limit at a characteristic excitation temperature of 10 - 50 K) rotational transitions of cyclopropenone, propynal, and propadienone between 4 GHz and 50 GHz were searched for in publicly available data from the PRIMOS survey of Sgr B2(N), a NRAO key project on the Robert C. Byrd Green Bank Telescope (GBT) from 2008 January through 2011 July\footnote{Access to the entire PRIMOS dataset and specifics on the observing strategy including the overall frequency coverage, is available at http://www.cv.nrao.edu/$\sim$aremijan/PRIMOS/}.  An LSR source velocity of +64 km s$^{-1}$ was assumed and antenna temperatures were recorded on the T$_A^*$ scale \citep{Ulich_&_Haas_1976} with estimated 20\% uncertainties.  Data were taken in position-switching mode, with two minute scans toward the on position ($\alpha$J2000 = 17$^{h}$47$^{m}$19.8$^{s}$, $\delta$J2000 = -28$\degree$22\arcmin17.0\arcsec) and towards the off position, 1$\degree$ east in azimuth.  Additional details of PRIMOS survey observations can be found in \cite{Neill_et_al_2012}.  Rest frequencies for all transitions were obtained from the Splatalogue database\footnote{Original laboratory data for PRIMOS frequency ranges is from \cite{Benson_et_al_1973, Winnewisser_1973, Brown_et_al_1987}.  Cyclopropenone and propadienone frequencies are catalogued through CDMS \citep{CDMS_2001, CDMS_2005},  and propynal through JPL \citep{JPL_1998}.}\footnote{Available at www.splatalogue.net \citep{Remijan_splatalogue_2007}.}.

\section{Results}

\subsection{Cyclopropenone}

Nine low-energy, high line-strength transitions of cyclopropenone were identified in absorption in PRIMOS data.  Of these, six had been previously identified by \cite{Hollis_et_al_2006} at lower sensitivity and spectral resolution (see Table 1).  All observed transitions were a-type ($\mu$$_{a}$ = 4.39(06) D) \citep{Benson_et_al_1973}, with six R-branch transitions ($\Delta$J = -1) and three Q-branch ($\Delta$J = 0).  Three higher energy (J \textgreater \space 4) Q-branch transitions were not observed.  This encompasses all strong transitions within the PRIMOS frequency coverage, with the exception of the 2$_{12}$ - 1$_{11}$ transition where observational data were not available.

All detected transitions were observed at a nominal rest velocity of v$_{LSR}$ = +64 km s$^{-1}$, with a second, weaker, v$_{LSR}$ = +82 km s$^{-1}$ component observed for all transitions other than 1$_{01}$ - 0$_{00}$;  consistent with other molecular detections with the GBT toward Sgr B2(N) (see, e.g. \cite{Hollis_et_al_2004, Remijan_et_al_2008}, and references therein).  Observations of all Q- and R-branch transition rest frequencies within the PRIMOS coverage are presented in Figures 2 and 3, respectively.  

\begin{figure*}[h!]
\centering
\includegraphics[scale=1]{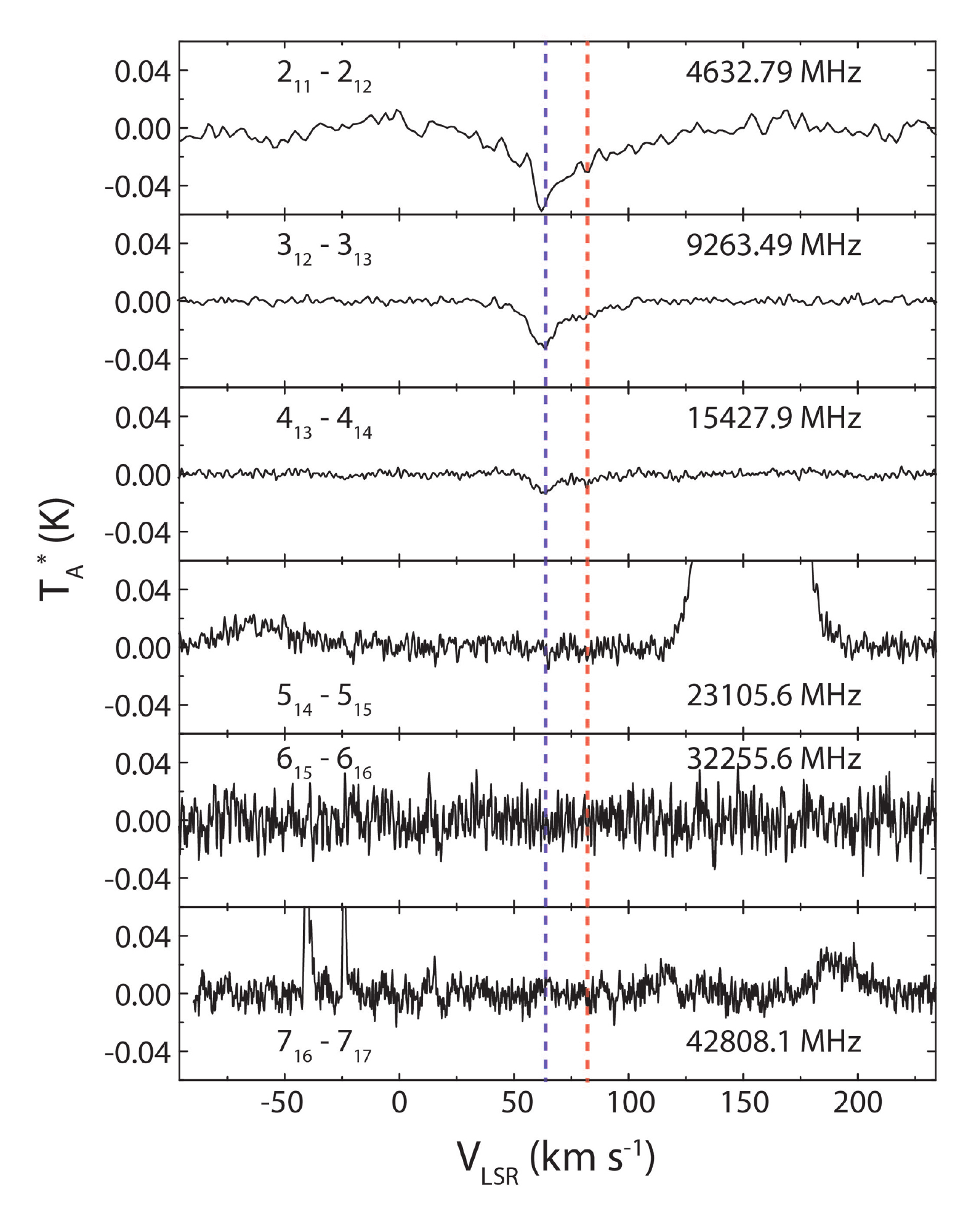}
\caption{{\small Observed cyclopropenone Q-branch transitions in PRIMOS data toward Sgr B2(N).  Dashed vertical lines indicate the primary +64 km s$^{-1}$ component (blue) and the secondary +82 km s$^{-1}$ component (red)}}
\end{figure*}

\begin{figure*}[t!]
\centering
\includegraphics[scale=0.8]{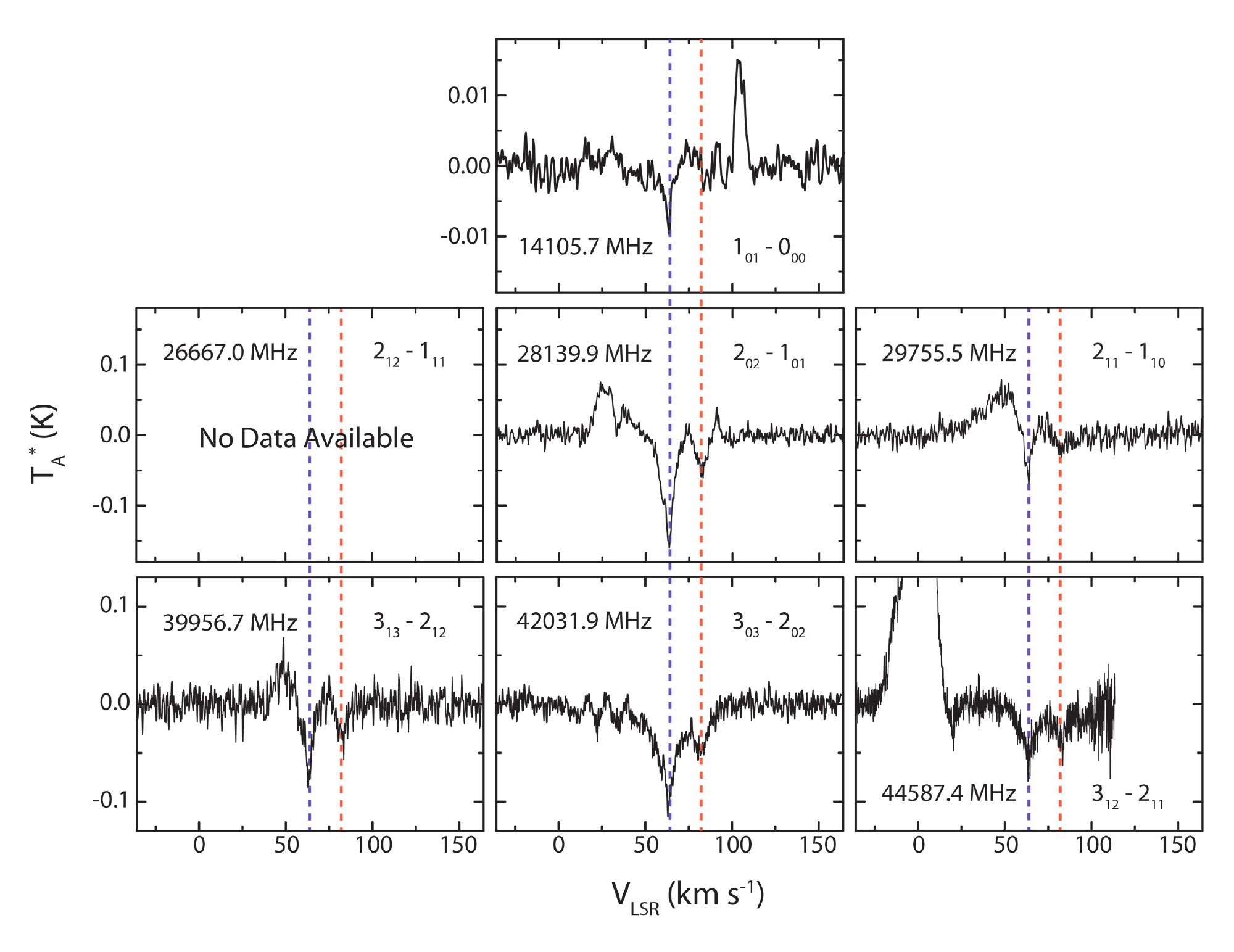}
\caption{{\small Observed cyclopropenone R-branch transitions in PRIMOS data toward Sgr B2(N).  Dashed vertical lines indicate the primary +64 km s$^{-1}$ component (blue) and the secondary +82 km s$^{-1}$ component (red)}}
\end{figure*}

As seen in Figure 3, the 2$_{11}$ - 1$_{10}$ transition is blended with a hydrogen recombination line (H(94)$\delta$) at 29575.189 MHz.  Additionally, the v$_{LSR}$ = +64 km s$^{-1}$ component of the 3$_{13}$ - 2$_{12}$ transition is partially blended with the 2$_{02}$ - 1$_{01}$, v=1 transition of NH$_{2}$CN in emission.  Best-fit Gaussian line widths and peak intensities or respective upper limits are reported in Table 1.  The 2$_{11}$ - 1$_{10}$ transition was simultaneously fit in absorption against a modeled Gaussian emission profile for the H(94)$\delta$ recombination line.

\begin{table*}
\begin{center}
\small
\begin{threeparttable}[b]
\caption{Observed transitions of cyclopropenone}
\begin{tabular}{+c^c^c^c^c^c^c^c}
\toprule
	Transition & Frequency$^{ab}$ & E$_{u}$ & S$_{ij}\mu^{2}$ & \multicolumn{2}{c} 64 km s$^{-1}$ & \multicolumn{2}{c} 82 km s$^{-1}$ \\
	J'$_{kk}$ - J"$_{kk}$ & (MHz) & (K) & (D$^{2}$) & $\Delta$T$_{A}^{*c}$ (mK) & $\Delta$V$^{c}$ (km s$^{-1}$) & $\Delta$T$_{A}^{*c}$ (mK) & $\Delta$V$^{c}$ (km s$^{-1}$) \\ 
	\otoprule
	2$_{11}$ - 2$_{12}$		&	4632.79	&	3.34		&	48.19	&	-57.2(5.4)			&	14.8(0.9)		&	-32.1(4.6)			&	15.1(1.2)		\\
	3$_{12}$ - 3$_{13}$$^d$	&	9263.49	&	5.48		&	33.76	&	-29.1(6.1)                	&	15.9(0.5)		&	-10.0(0.9)			&	12.4(1.1)		\\
	4$_{13}$ - 4$_{14}$$^d$	&	15427.90	&	8.33		&	26.12	&	-12.1(0.6)			& 	11.6(0.6)		&	-5.4(0.7)			&	12.0(0.8)		\\
	5$_{14}$ - 5$_{15}$		&	23105.58	&	11.88	&	21.41	&	$<$ $\space$5.0	&	-			& 	$<$ $\space$5.0	&	-			\\
	6$_{15}$ - 6$_{16}$		&	32255.63	&	16.14	&	18.27	&	$<$ $\space$10.7	&	-			& 	$<$ $\space$10.7	&	-			\\
	7$_{16}$ - 7$_{17}$		&	42808.10	&	21.08	&	16.10	&	$<$ $\space$7.3	&	-			& 	$<$ $\space$7.3	&	-			\\
	                                		&			&			&	 		&	    				&				&					&				\\
	1$_{01}$ - 0$_{00}$$^d$	&	14105.74	&	0.68		&	19.27	&	-6.7(0.7)			&	7.2(0.9)		& 	$<$ $\space$2.2	&	-			\\
	2$_{12}$ - 1$_{11}$		&	26667.02	&	3.12		&	86.73	&	\nodata			&	-			& 	\nodata			&	-			\\
	2$_{02}$ - 1$_{01}$		&	28139.89	&	2.03		&	38.52	&	-133.6(4.2)		&	8.6(0.3)		& 	-48.9(4.5)			&	6.4(0.5)		\\
	2$_{11}$ - 1$_{10}$		&	29755.51	&	3.34		&	86.73	&	-83.2(8.8)			&	8.2(0.5)		&	-22.9(7.3)			&	14.2(2.4)		\\
	3$_{13}$ - 2$_{12}$$^d$	&	39956.70	&	5.04		&	154.15	&	-66.5(3.1)    		&	6.0(0.3)		&	-32.9(3.0)			&	5.9(0.6)		\\
	3$_{03}$ - 2$_{02}$$^d$	&	42031.94	&	4.04		&	57.71	&	-74.5(1.9)    		&	15.3(0.4)		& 	-40.3(2.2)			&	9.4(0.5)		\\ 
	3$_{12}$ - 2$_{11}$$^d$	&	44587.40	&	5.48		&	154.1	&	-46.9(2.1)			&	11.7(0.6)		& 	-34.4(1.8)			&	12.5(0.9)		\\ 
	\bottomrule
\end{tabular}
\begin{tablenotes}
\item[a] Beam sizes, efficiencies, and continuum temperatures for each respective frequency can be found in \cite{Hollis_et_al_2007}
\item[b] Rest frequencies are all taken from Splatalogue, see Section 2 for complete references
\item[c] The uncertainties for the intensities and line widths are Type B, $k$ = 1 (1$\sigma$) \citep{Taylor_&_Kuyatt_1994}
\item[d] Previously detected with the GBT with lower sensitivity and spectral resolution \citep{Hollis_et_al_2006}
\end{tablenotes}
\end{threeparttable}
\end{center}
\end{table*}

\subsection{Propynal}

Six uncontaminated, low-energy, high line-strength transitions of propynal were identified in absorption.  One of these had been previously been detected by \cite{Hollis_et_al_2004_propynal} with lower sensitivity and spectral resolution (see Table 2).  Three of the identified transitions were k$_{a}$ = 0 a-type ($\mu$$_{a}$ = 2.359(18)  D), and three were b-type ($\mu$$_{b}$ = 1.48(22) D) \citep{Brown_&_Godfrey_1984}. Of the remaining twelve low-energy, high line-strength transitions predicted to be possibly observable, six were not observed above the noise level, five were contaminated by blends with other molecule emission features (see Figure 4), and frequency coverage was not available for one transition.  Notably, although multiple k$_{a}$ = 0 a-type transitions were observed with strong absorption, no k$_{a}$ = 1 a-type transitions were detected.

\cite{Irvine_et_al_1988} suggested that the k$_{a}$ = 1 transitions of propynal may be weaker than predicted by a thermal model due to b-type transitions transferring k$_{a}$ = 1 population into the k$_{a}$ = 0 state; this behavior has also been observed in HNCO \citep{Hocking_et_al_1974}.  Irvine et al. used a collisional model to predict corrected k$_{a}$ = 1 intensities for their detection in TMC-1, and determined an approximate intensity in agreement with their tentative 2 - 3$\sigma$ detection of the 2$_{11}$ - 1$_{10}$ transition using the NRAO 140 ft.  If this model is correct, a rough scaling to our observations would explain the lack of k$_{a}$ = 1 detections, with all predicted intensities below the observed noise level.

All detected transitions were observed at a nominal rest velocity of v$_{LSR}$ = +64 km s$^{-1}$, with an additional, weaker, v$_{LSR}$ = +82 km s$^{-1}$ component observed for all transitions other than 4$_{04}$ - 3$_{03}$, where the noise level is comparable to the expected intensity of the  v$_{LSR}$ = +82 km s$^{-1}$ component.  Observations of all a-type and b-type transition rest frequencies within PRIMOS coverage are presented in Figures 4 and 5, respectively.  Best-fit Gaussian line widths and peak intensities or respective upper limits are reported in Table 2.  All features were observed with line widths of $\sim$4 - 10 km s$^{-1}$ with the exception of the 5$_{15}$ - 6$_{06}$ transition, where the velocity components were blended due to low S/N and line widths were therefore fixed together during fitting, possibly artificially broadening the fit width.

\begin{figure*}[h!]
\centering
\includegraphics[scale=1]{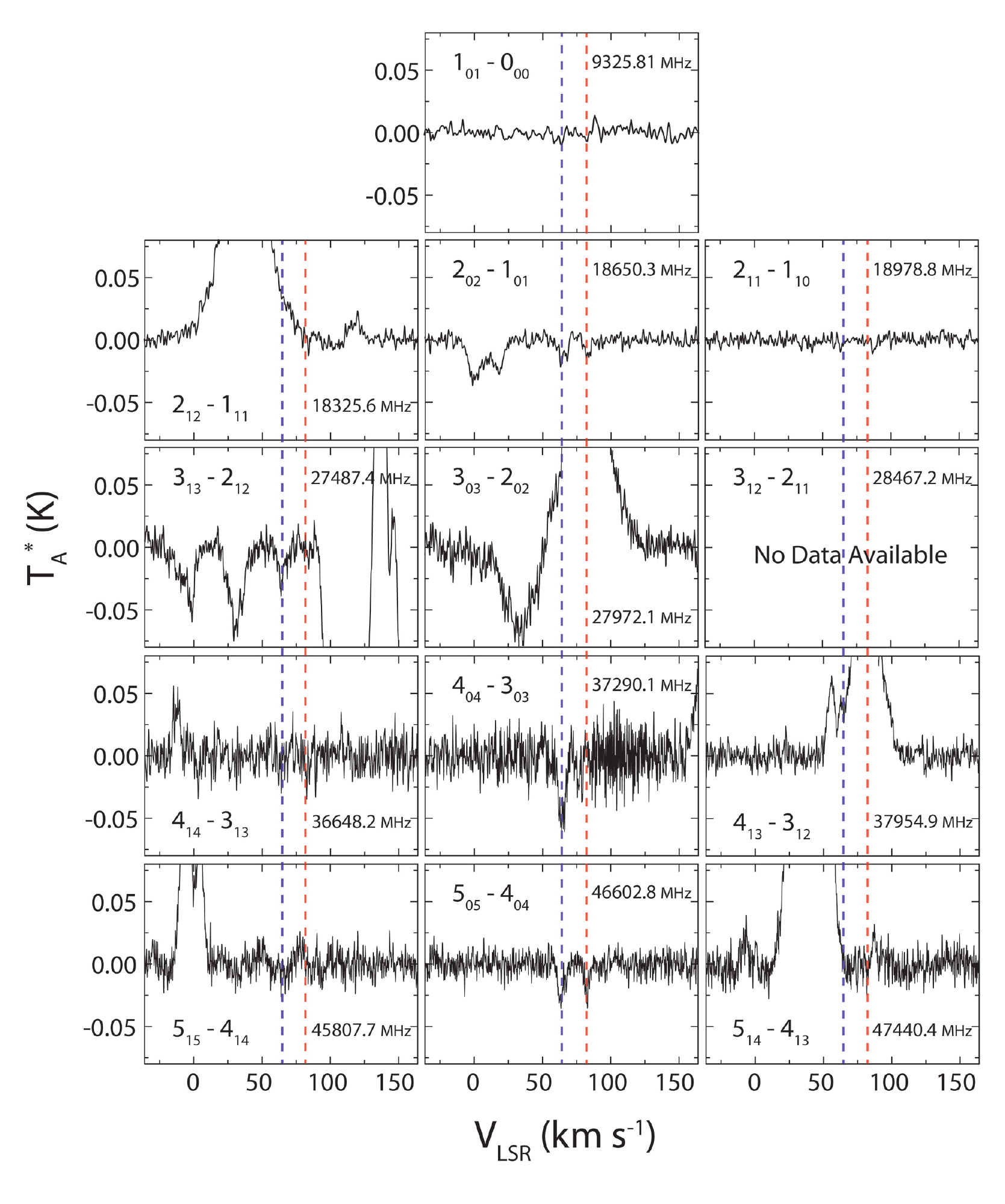}
\caption{{\small Observed propynal a-type transitions in PRIMOS data toward Sgr B2(N).  Dashed vertical lines indicate the primary +64 km s$^{-1}$ component (blue) and the secondary +82 km s$^{-1}$ component (red)}}
\end{figure*}

\begin{figure*}[h!]
\centering
\includegraphics[scale=1]{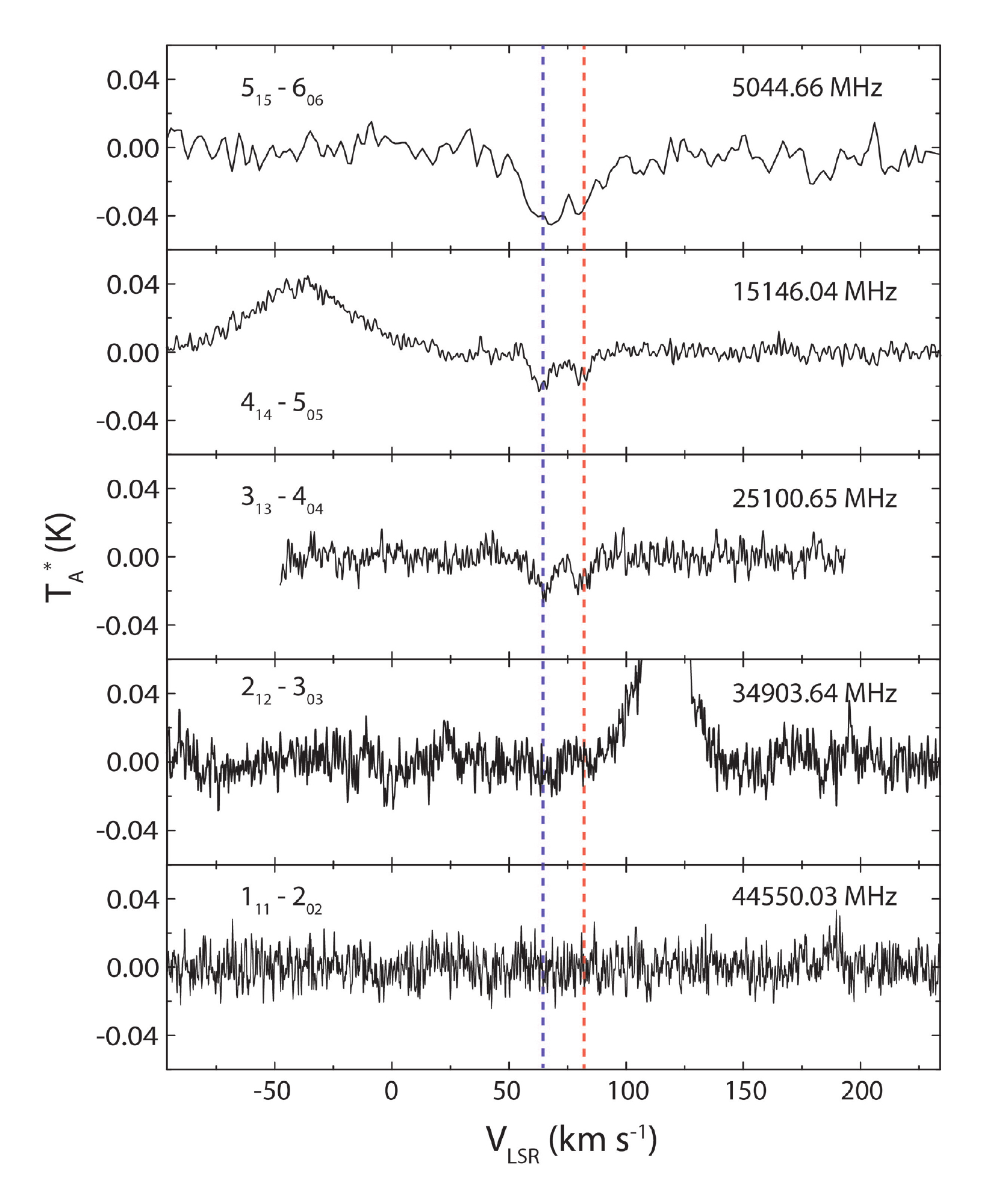}
\caption{{\small Observed propynal b-type transitions in PRIMOS data toward Sgr B2(N).  Dashed vertical lines indicate the primary +64 km s$^{-1}$ component (blue) and the secondary +82 km s$^{-1}$ component (red)}}
\end{figure*}

\begin{table*}
\begin{center}
\small
\begin{threeparttable}[b]
\caption{Observed transitions of propynal}
\begin{tabular}{+c^c^c^c^c^c^c^c}
\toprule
	Transition & Frequency$^{ab}$ & E$_{u}$ & S$_{ij}\mu^{2}$ & \multicolumn{2}{c} 64 km s$^{-1}$ & \multicolumn{2}{c} 82 km s$^{-1}$ \\
	J'$_{kk}$ - J"$_{kk}$ & (MHz) & (K) & (D$^{2}$) & $\Delta$T$_{A}^{*c}$ (mK) & $\Delta$V$^{c}$ (km s$^{-1}$) & $\Delta$T$_{A}^{*c}$ (mK) & $\Delta$V$^{c}$ (km s$^{-1}$) \\ 
	\otoprule
	5$_{15}$ - 6$_{06}$		&	5044.66  	&	9.64	&	5.58  	&	-36.4(2.2)            	&	20.3(1.8)$^{d}$	&	-26.1(2.3) 			&	20.3(1.8)$^{d}$	\\
	4$_{14}$ - 5$_{05}$$^e$	&	15146.04	&	7.44	&	4.43  	&	-20.3(1.2)            	&	9.9(0.6)		&	-14.5(1.3)			&	10.3(0.9)		\\
	3$_{13}$ - 4$_{04}$		&	25100.65	&	5.68	&	3.30  	&	-20.4(1.5) 			& 	10.1(0.8)   	&	-16.6(1.6) 			&	8.3(0.9)		\\
	2$_{12}$ - 3$_{03}$		&	34903.64	&	4.36	&	2.18  	&	$<$ $\space$8.7	&	-			& 	$<$ $\space$8.7	&	-			\\
	1$_{11}$ - 2$_{02}$		&	44550.03	&	3.48	&	1.09  	&	$<$ $\space$8.6	&	-			& 	$<$ $\space$8.6 	&	-			\\
	                                		&	               	&	       	&	         	&	                            	&				&					&				\\
	1$_{01}$ - 0$_{00}$		&	9325.81  	&	0.45	&	5.56  	&	$<$ $\space$3.8	&	-			& 	$<$ $\space$3.8 	&	-			\\
	2$_{12}$ - 1$_{11}$		&	18325.56	&	4.36	&	8.35  	&	Blend                  	&	-			& 	Blend 			&	-			\\
	2$_{02}$ - 1$_{01}$		&	18650.31	&	1.34	&	11.13	&	-16.8(1.6)            	&	7.2(0.7)		& 	-14.1(1.5) 			&	4.9(0.6)		\\
	2$_{11}$ - 1$_{10}$		&	18978.79	&	4.41	&	8.35  	&	$<$ $\space$3.6	&	- 			&	$<$ $\space$3.6 	&	-			\\
	3$_{13}$ - 2$_{12}$		&	27487.45	&	5.68	&	14.84	&	Blend                  	&	- 			&	$<$ $\space$8.3 	&	-			\\
	3$_{03}$ - 2$_{02}$		&	27972.13	&	2.68	&	16.69	&	Blend                  	&	- 			& 	Blend 			&	-			\\ 
	3$_{12}$ - 2$_{11}$		&	28467.15	&	5.77	&	14.84	&	\nodata              	&	- 			& 	\nodata 		&	-			\\ 
	4$_{14}$ - 3$_{13}$		&	36648.20	&	7.44	&	20.87	&	$<$ $\space$11.2	&	- 			& 	$<$ $\space$11.2 	&	-			\\ 
	4$_{04}$ - 3$_{03}$		&	37290.10	&	4.47	&	22.26	&	-52.7(-4.1)           	&	5.8(0.5) 		& 	$<$ $\space$16.4 	&	-			\\ 
	4$_{13}$ - 3$_{12}$		&	37954.95	&	7.60	&	20.87	&	Blend                  	&	- 			& 	Blend             		&	-			\\ 
	5$_{15}$ - 4$_{14}$		&	45807.70	&	9.64	&	26.71	&	$<$ $\space$9.5 	&	- 			& 	$<$ $\space$9.5 	&	-			\\ 
	5$_{05}$ - 4$_{04}$		&	46602.80	&	6.71	&	27.82	&	-28.8(2.1)            	&	5.6(0.3) 		& 	-23.4(1.7) 			&	4.0(0.4)		\\ 
	5$_{14}$ - 4$_{13}$		&	47440.40	&	9.87	&	26.71	&	Blend                  	&	- 			& 	$<$ $\space$9.2 	&	-			\\
	\bottomrule
\end{tabular}
\begin{tablenotes}
\item[a] Beam sizes, efficiencies, and continuum temperatures for each respective frequency can be found in \cite{Hollis_et_al_2007}
\item[b] Rest frequencies are all taken from Splatalogue, see Section 2 for complete references
\item[c] The uncertainties for the intensities and line widths are Type B, \emph{k} = 1 (1$\sigma$) \citep{Taylor_&_Kuyatt_1994}
\item[d] Due to blended velocity components, line widths were shared during fitting
\item[e] Previously detected with the GBT with lower sensitivity and spectral resolution \citep{Hollis_et_al_2004_propynal}
\end{tablenotes}
\end{threeparttable}
\end{center}
\end{table*}

\subsection{Propadienone}

No transitions of propadienone were identified in PRIMOS data.  Of the known transitions between 4 GHz and 50 GHz, only a-type transitions ($\mu$$_{a}$ = 2.156 D, $\mu$$_{b}$ = 0.7914 D) \citep{Brown_et_al_1981} are predicted to be strong enough to be observed in PRIMOS data. It is important to note that the slight bend of the C$_{3}$O chain in propadienone causes a small barrier, creating two tunneling states separated by a small (0.1198 cm$^{-€"1}$) energy difference.  A-type transitions occur within each substate while b-type transitions occur between the two tunneling states \citep{Brown_et_al_1987}.  B-type transitions within PRIMOS frequency coverage are predicted to be significantly weaker due to a lower dipole moment and high upper state energies.  Observations of all a-type transition rest frequencies within the PRIMOS coverage are presented in Figure 6 and upper limits on all a-type transition intensities are given in Table 3.

\begin{figure*}[h!]
\centering
\includegraphics[scale=0.8]{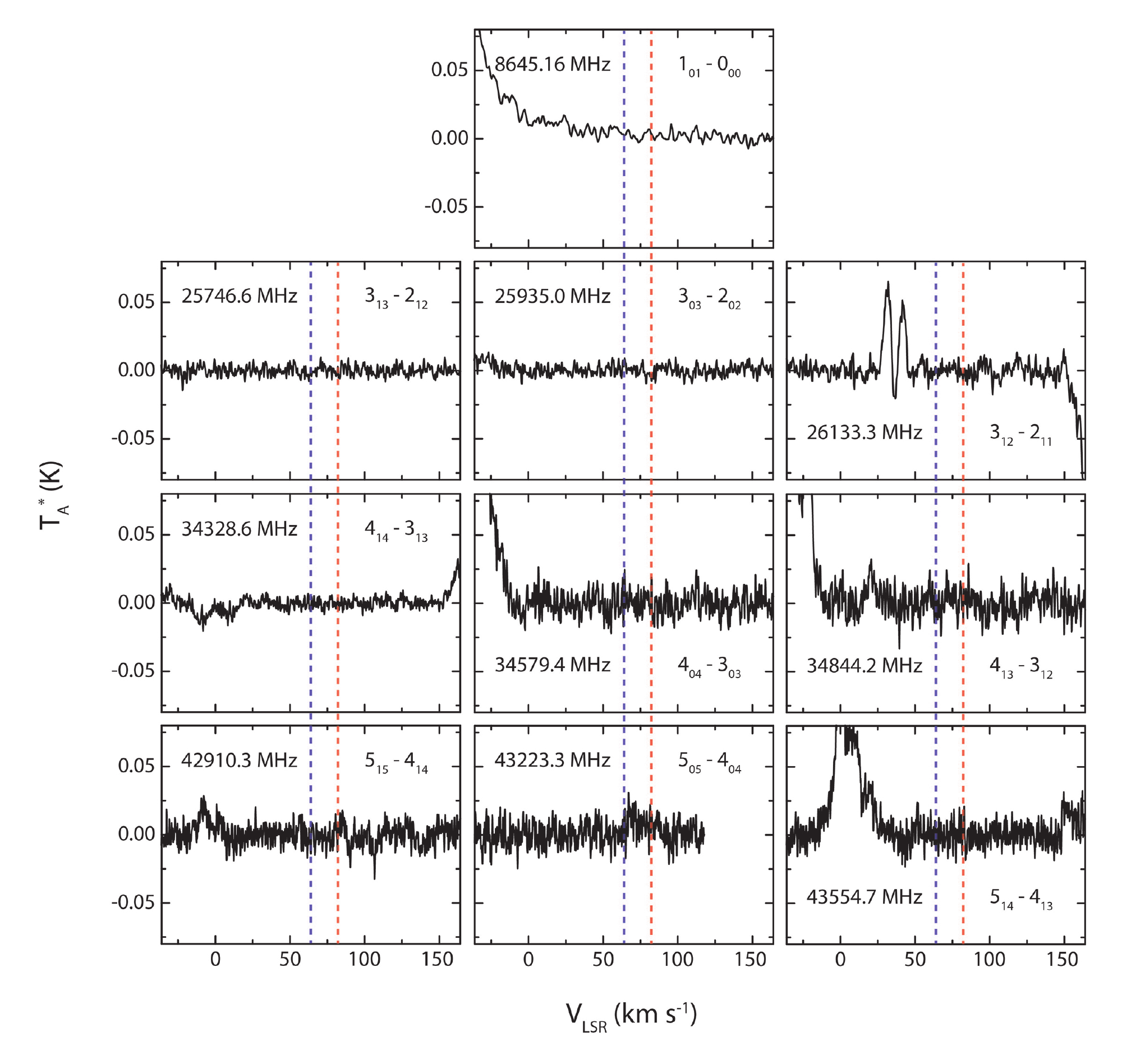}
\caption{{\small Observed propadienone transitions in PRIMOS data toward Sgr B2(N).  Dashed vertical lines indicate the primary +64 km s$^{-1}$ component (blue) and the secondary +82 km s$^{-1}$ component (red)}}
\end{figure*}

\begin{table*}
\begin{center}
\small
\begin{threeparttable}[b]
\caption{Observed transitions of propadienone}
\begin{tabular}{+c^c^c^c^c^c^c^c}
\toprule
	Transition & Frequency$^{ab}$ & E$_{u}$ & S$_{ij}\mu^{2}$ & \multicolumn{2}{c} 64 km s$^{-1}$ & \multicolumn{2}{c} 82 km s$^{-1}$ \\
	J'$_{kk}$ - J"$_{kk}$ & (MHz) & (K) & (D$^{2}$) & $\Delta$T$_{A}^{*c}$ (mK) & $\Delta$V$^{c}$ (km s$^{-1}$) & $\Delta$T$_{A}^{*c}$ (mK) & $\Delta$V$^{c}$ (km s$^{-1}$) \\ 
	\otoprule
	1$_{01}$ - 0$_{00}$	&	8645.16	&	0.41		&	4.65		&	$<$ $\space$3.4	&	-	& 	$<$ $\space$3.4	&	-	\\
	2$_{12}$ - 1$_{11}$	&	17164.49	&	8.39		&	20.92	&	No Data			&	-	& 	No Data			&	-	\\
	2$_{02}$ - 1$_{01}$	&	17290.18	&	1.24		&	9.30		&	No Data			&	-	& 	No Data			&	-	\\
	2$_{11}$ - 1$_{10}$	&	17422.33	&	8.41		&	20.92	&	No Data			&	- 	&	No Data			&	-	\\
	3$_{13}$ - 2$_{12}$	&	25746.59	&	9.63		&	37.19	&	$<$ $\space$3.4	&	- 	&	$<$ $\space$3.4	&	-	\\
	3$_{03}$ - 2$_{02}$	&	25934.97	&	2.49		&	13.95	&	$<$ $\space$3.6	&	- 	& 	$<$ $\space$3.6	&	-	\\ 
	3$_{12}$ - 2$_{11}$	&	26133.34	&	9.66		&	37.19	&	$<$ $\space$5.4	&	- 	& 	$<$ $\space$5.4	&	-	\\ 
	4$_{14}$ - 3$_{13}$	&	34328.55	&	11.27	&	52.29	&	$<$ $\space$3.3	&	- 	& 	$<$ $\space$3.3	&	-	\\ 
	4$_{04}$ - 3$_{03}$	&	34579.38	&	4.15		&	18.60	&	$<$ $\space$8.0	&	-	& 	$<$ $\space$8.0	&	-	\\ 
	4$_{13}$ - 3$_{12}$	&	34844.16	&	11.34	&	52.29	&	$<$ $\space$9.2	&	- 	& 	$<$ $\space$9.2	&	-	\\ 
	5$_{15}$ - 4$_{14}$	&	42910.28	&	13.33	&	66.93	&	$<$ $\space$6.9	&	- 	& 	$<$ $\space$6.9	&	-	\\ 
	5$_{05}$ - 4$_{04}$	&	43223.30	&	6.22		&	23.24	&	$<$ $\space$8.6	&	-	& 	$<$ $\space$8.6	&	-	\\ 
	5$_{14}$ - 4$_{13}$	&	43554.72	&	13.43	&	66.94	&	$<$ $\space$7.6	&	- 	& 	$<$ $\space$7.6	&	-	\\
	\bottomrule
\end{tabular}
\begin{tablenotes}
\item[a] Beam sizes, efficiencies, and continuum temperatures for each respective frequency can be found in \cite{Hollis_et_al_2007}
\item[b] Rest frequencies are all taken from Splatalogue, see Section 2 for complete references
\item[c] The uncertainties for the intensities and line widths are Type B, \emph{k} = 1 (1$\sigma$) \cite{Taylor_&_Kuyatt_1994}
\end{tablenotes}
\end{threeparttable}
\end{center}
\end{table*}

\section{Discusion}

Although our observational results support a detection of cyclopropenone and propynal and a non-detection of propadienone, it is important to additionally examine their abundance ratios.  It may be possible for a given isomer to be more abundant, yet not detected, due to inherent line strengths or excitation issues.  Using the formulation in \cite{Hollis_et_al_2004}, and assuming a temperature of 15K, similar to other cold large organics detected with the PRIMOS survey, we calculate column densities of $\sim$10$^{12}$ cm$^{-2}$ for cyclopropenone, $\sim$10$^{13}$ cm$^{-2}$ for propynal, and an upper limit of $<$10$^{11}$ cm$^{-2}$ for propadienone.  Thus, if propadienone was actually equal in abundance to cyclopropenone or propynal, we would expect to see features 10-100 times the noise level, yet no such features are observed. These calculations assume continuum temperatures for Sgr B2(N) from \cite{Hollis_et_al_2007}, as well as LTE conditions.  We wish to stress that our observations are not fit particularly well with these assumptions, suggesting that a large number of the observed lines are sub-thermally excited.  If these features arise from non-LTE conditions, this may call into question our derived column densities, but would also invalidate the assumptions necessary for the MEP, strengthening our assertion that thermodynamically driven processes do not control the bulk of interstellar abundance ratios.

These results answer calls for further observational tests of the MEP from theoretical studies investigating the relative energies of multiple isomer families \citep{Lovas_et_al_2010, Karton_&_Talbi_2014}.  In particular, \cite{Karton_&_Talbi_2014} explicitly note that within the C$_3$H$_2$O isomer family, propadienone should be detected based on the predictions of the MEP and suggest further observational searches for the molecule.  Its non-detection now leaves us with the the question as to which synthesis routes result in the formation of cyclopropenone and propynal but not propadienone.

\subsection{Formation chemistry on grain surfaces}

In the interstellar medium, dust grains serve as ``catalysts'' for much of the rich organic chemistry that is observed \citep{Herbst_&_van_Dishoeck_2009}. \cite{Zhou_et_al_2008} studied the non-equilibrium grain-ice chemistry in cold ($\sim$ 10 K) regions of the ISM, finding that cyclopropenone is formed on grain surfaces via reactions between triplet CO and C$_2$H$_2$, i.e.

\begin{equation}
 \mathrm{CO} + \mathrm{C_2H_2} \rightarrow \mathrm{c-C_3H_2O},
\end{equation}

\noindent and that propynal is formed in the bulk by carbon monoxide and acetylene through a radical pair reaction, i.e.

\begin{equation}
\mathrm{CO} + \mathrm{C_2H_2} \rightarrow [\mathrm{HCO-CCH}]^* \rightarrow \mathrm{HCCHO}.
\end{equation}

\cite{Zhou_et_al_2008} found that cosmic ray protons ($\sim$10 MeV) are likely the source of energy that drives these reactions, inducing a C-H cleavage in collisions with acetylene that initiates reactions (1) and (2). Experimental results from ultra-high vacuum irradiation of ice-samples showed that the two metastable isomers of C$_3$H$_2$O were produced, with only a tentative detection of propadienone. The molecules formed in this way can then enrich the gas-phase abundance through sublimation, a particularly likely scenario in star-forming regions such as Sgr B2(N).

\subsection{Gas-phase formation chemistry}

In addition to formation pathways involving grain chemistry, there are several gas-phase reactions that are thought to produce cyclopropenone and propynal, but no feasible reactions have been proposed for propadienone production \citep{Kwon_et_al_2006, Petrie_1995, Irvine_et_al_1988}. A study by \cite{Quan_&_Herbst_2007} attempted, in part, to reproduce the observed abundance of c-C$_3$H$_2$O in Sgr B2(N). Their network included the gas-phase formation pathway

\begin{equation}
\mathrm{C_3H^+} + \mathrm{H_2} \rightarrow \mathrm{c-C_3H_3^+} + h\nu
\end{equation}

\begin{equation}
\mathrm{c-C_3H_3^+} + e^- \rightarrow \mathrm{c-C_3H_2} + \mathrm{H}
\end{equation}

\begin{equation}
\mathrm{c-C_3H_2} + \mathrm{O}(^3P) \rightarrow \mathrm{c-C_3H_2O}.
\end{equation}

Recently, the reactant C$_3$H$^+$ in reaction (3) was detected in Sgr B2(N), suggesting the validity of this gas-phase formation route as a possible source of C$_3$H$_2$ \citep{McGuire_et_al_2013}.\footnote{Interested readers should note that new work on the formation mechanisms of cyclopropenone \citep{Ahmadvand_et_al_2014} was published very late in the review process for this paper.  Ahmadvand et al. claim that the gas-phase formation pathway suggested by \cite{Quan_&_Herbst_2007} is spin-forbidden, while the grain surface pathway in Section 4.1 is spin-allowed.}  The abundance of the cyclic form of C$_3$H$_2$ has been estimated to be 50-150 times greater than the linear form of C$_3$H$_2$ toward Sgr B2(N) \citep{Cernicharo_et_al_1999}, suggesting that it is present in abundance and has the potential to be a parent molecule of c-C$_3$H$_2$O \citep{Matthews_&_Irvine_1985}.  The model of \cite{Quan_&_Herbst_2007} was unable to reproduce the observed abundance of c-C$_3$H$_2$O in Sgr B2(N), but this is not too surprising, given that they did not include grain-surface reactions for the formation of cyclopropenone in their network.

\subsection{Interpretation of this case study and conclusions}

From this collection of reactions, a picture emerges in Sgr B2(N) that there are simply more non-equilibrium formation pathways for propynal and cyclopropenone (gas and grain) than for propadienone, despite the fact that it is thermodynamically more stable. Additionally, \cite{Lovas_et_al_2010} suggest that in grain-surface reactions, the surrounding ice matrix may quench excess energy, preventing rearrangement to more thermodynamically favorable species such as propadienone.  These results are not limited to our current work, however; previous studies of other isomer systems such as HNC/HCN have also suggested the importance of kinetic control in determining abundance ratios.

The ground state energy of the metastable HNC is 7240 K above HCN (Bowman et al 1993). Thus, if \cite{Lattelais_et_al_2009} were correct in stating that relative isomeric abundances can be explained in terms of ground state energies, one would expect HCN to be much more abundant than HNC. However, observations by \cite{Schilke_et_al_1992} of OMC-1 showed an HNC/HCN ratio of almost unity. Further observations \citep{Padovani_et_al_2011, Hebrard_et_al_2012, Aalto_et_al_2012} demonstrate that as temperatures decrease, the HNC/HCN ratio increases. Studies by \cite{Herbst_et_al_2000} and \cite{Graninger_et_al_2014} indicate that the relative abundances of HCN and HNC can best be understood in terms of kinetic control, specifically that certain destruction pathways for HNC involving neutral-neutral reactions with H do not occur at low temperatures due to energy barriers.

We therefore conclude that the minimum energy principle does not have substantial predictive power for determining the most likely isomer to be observed in an given isomer family, especially in blind application with no prior knowledge. This suggests that as next generation capabilities become available, specific emphasis should be placed on investigating molecules and molecular reactions under kinetic control, especially when investigating regions where LTE assumptions are unlikely to hold.  With a better understanding of chemical environments, formation and destruction mechanisms, and excitation conditions, it may be possible to apply a modified version of the MEP for certain isomer families in specific environments, but further observational and laboratory work is necessary for rigorous predictions.


\acknowledgments

R.A.L. gratefully acknowledges funding from an NSF Graduate Research Fellowship and the Virginia Space Grant Program, as well as support from the College Science Scholars program at the University of Virginia.  B.A.M. gratefully acknowledges funding from an NSF Graduate Research Fellowship and G.A. Blake for his support.  We thank an anonymous referee for important comments.  The National Radio Astronomy Observatory is a facility of the National Science Foundation operated under cooperative agreement by Associated Universities, Inc.




\bibliographystyle{apj}
\bibliography{refs}

%% The following command ends your manuscript. LaTeX will ignore any text

\end{document}